\documentclass[a4paper,pra,showpacs,superscriptaddress,twocolumn]{revtex4-1}

\usepackage{hyperref}
\usepackage{amsfonts,dsfont,mathrsfs,amsmath,amssymb,bm,bbm}
\usepackage[english]{babel}
\usepackage[babel]{microtype}  
\usepackage{graphicx}
\usepackage{subfigure}
\usepackage{braket}
\usepackage{color}

\newcommand{\myref}[1]{Eq.\hspace{0.3em}\eqref{#1}}
\newcommand{\mycite}[1]{Ref.\hspace{0.3em}\cite{#1}}
\newcommand{\mycites}[1]{Refs.\hspace{0.3em}\cite{#1}}
\newcommand{\myfig}[1]{Fig.\hspace{0.3em}\ref{#1}}
\newcommand{\mysec}[1]{Sec.\hspace{0.3em}\ref{#1}}
\newcommand{\tred}{\color{black}}

\newcommand{\mc}[1]{\mathcal{#1}}

\newcommand{\bgeq}{\begin{equation}}
	\newcommand{\edeq}{\end{equation}}
\newcommand{\bgal}{\begin{align}}
	\newcommand{\edal}{\end{align}}
\newcommand{\bgald}{\begin{aligned}}
	\newcommand{\edald}{\end{aligned}}
\newcommand{\bgfg}{\begin{figure}}
	\newcommand{\edfg}{\end{figure}}

\begin{document}

	\title{Entropic uncertainty relations for multiple measurements assigned with biased weights}
	
	\author{Shan Huang}
	\affiliation{National Laboratory of Solid State Microstructures and School of Physics,  Collaborative Innovation Center of Advanced Microstructures, Nanjing University, Nanjing 210093, China}
	\affiliation{Institute for Brain Sciences and Kuang Yaming Honors School, Nanjing University, Nanjing 210023,
		China}
    \affiliation{Hefei National Laboratory, University of Science and Technology of China, Hefei 230088, China}

	\author{Hua-Lei Yin}
	
	\email{hlyin@ruc.edu.cn}
	\affiliation{Department of Physics and Beijing Key Laboratory of Opto-Electronic Functional Materials and Micro-Nano Devices, Key Laboratory of Quantum State Construction and Manipulation (Ministry of Education), Renmin University of China, Beijing 100872, China}
	\affiliation{National Laboratory of Solid State Microstructures and School of Physics, Collaborative Innovation Center of Advanced Microstructures, Nanjing University, Nanjing 210093, China}
	
	\author{Zeng-Bing Chen}
	\email{zbchen@nju.edu.cn}
	\affiliation{National Laboratory of Solid State Microstructures and School of Physics, Collaborative Innovation Center of Advanced Microstructures, Nanjing University, Nanjing 210093, China}
	
	\author{Shengjun Wu}
	\email{sjwu@nju.edu.cn}
	\affiliation{National Laboratory of Solid State Microstructures and School of Physics, Collaborative Innovation Center of Advanced Microstructures, Nanjing University, Nanjing 210093, China}
	\affiliation{Institute for Brain Sciences and Kuang Yaming Honors School, Nanjing University, Nanjing 210023,
		China}
    \affiliation{Hefei National Laboratory, University of Science and Technology of China, Hefei 230088, China}

	\begin{abstract}
The entropic way of formulating Heisenberg's uncertainty principle not only plays a fundamental role in applications of quantum information theory but also is essential for manifesting genuine nonclassical features of quantum systems. In this paper we investigate R\'{e}nyi entropic uncertainty relations (EURs) in the scenario where measurements on individual copies of a quantum system are selected with nonuniform probabilities. In contrast with  EURs that characterize an observer's overall lack of information about outcomes with respect to a collection of measurements, we establish state-dependent lower bounds on the weighted sum of entropies over multiple measurements. Conventional EURs thus correspond to the special cases when all weights are equal, and in such cases, we show our results are generally stronger than previous ones. Moreover, taking the entropic steering criterion as an example, we numerically verify that our EURs could be advantageous in practical quantum tasks by optimizing the weights assigned to different measurements. Importantly, this optimization does not require quantum resources and is efficiently computable on classical computers.
	\end{abstract}
	
	\maketitle
	
	\section{Introduction}
Heisenberg's uncertainty principle \cite{heisenberg1927} is a fundamental concept in quantum mechanics which underlies one of the most important nonclassical features of quantum physics---quantum observables can be incompatible such that no observer has precise knowledge about them simultaneously. As a consequence, an observer's ability to predict (or certainty about) outcomes of measuring incompatible observables is inherently limited.

Various uncertainty measures have been proposed to formulate the uncertainty principle quantitatively. Among them, entropies are recognized as natural uncertainty measures from an information-theoretical perspective \cite{deutsch1983,maassen1988,wehner2010,bialynicki2011,coles2012,friedland2013,dammeier2015}. Entropic uncertainty relations (EURs) set lower bounds on one's uncertainty (lack of information) about measurement outcomes and, thus, are basic tools for the security analysis of quantum protocols, including quantum key distribution \cite{tomamichel2011,tomamichel2012,cerf2002,grosshans2004,berta2010,tan2021,xie2022} and quantum random number generation \cite{marangon2017,drahi2020,mazzucchi2021,liu2023} in the finite key scenario.  EURs can also be utilized to demonstrate genuine nonclassical features of quantum systems. For example, the uncertainty (unpredictability) of local measurement outcomes is a signature of weak classical correlation, yet quantum correlation is exclusive with local certainty \cite{yan2013}. This endows EURs with special significance in witnessing entanglement \cite{braunstein1988,giovannetti2004,guhne2004,guhneOLM2004,huang2010,horodecki1996,kurzynski2012}. (See the review  \cite{coles2017} for more applications.)

Following Deutsch \cite{deutsch1983} and the conjecture of Kraus \cite{kraus1987}, Maassen and Uffink \cite{maassen1988} proved the famous EUR for two non-degenerate observables:
\begin{align}
	H(\mc{M}_1)_\rho+H(\mc{M}_2)_\rho\geq -\log c_{\rm max}=: q_{\rm MU},       \label{qmu}
\end{align}
where $H(\mc{M}_\theta)_\rho=-\sum_ip_{i|\theta}\log p_{i|\theta}$ (all logarithms are base 2 in this work) is the Shannon entropy of the probability distribution $(p_{0|\theta},p_{1|\theta},\cdots)$ induced by measuring the $\theta$th observable on the quantum state $\rho$, and $c_{\rm max}=\max_ {i,j}\{\lvert\braket{i_1|j_2}\rvert^2\}$ denotes the maximal overlap between the normalized eigenbases $\{\ket{i_1}\}$ and $\{\ket{j_2}\}$ of observables under consideration, which characterizes the measurement incompatibility.  Notably, the entropic lower bound $q_{\rm MU}$ is nontrivial ($q_{\rm MU}>0$) for observables with no common eigenstate ($c_{\rm max}<1$). Furthermore, for two mutually unbiased bases (MUBs) of $d$-dimensional Hilbert space $\mc{H}_d$, i.e.,  $|\braket{i_1|j_2}|^2=1/d$  for all $i,j=0,1,\cdots,d-1$, $q_{\rm MU}=\log d$ is maximal and tight.

Attempts \cite{liu2015, xie2021} have been made to generalize \myref{qmu} to multiple measurement bases, in terms of the overlap between bases. When considering multiple measurements, however, merely the maximal overlap could be too rough a characterization of measurement incompatibility to ensure strong enough EURs. {\tred Partovi introduced the idea of utilizing majorization relations to investigate uncertainty relations \cite{partovi2011}. Inspired by this, Friedland \emph{et al} \cite{friedland2013} proposed a universal method of explicitly formulating EURs from the measurements under consideration. Further in-depth study shows that the majorization EURs, particularly the ones obtained by Rudinicki \emph{et al} \cite{puchala2013, rudnicki2014}, are generally stronger than other existing EURs except for multiple bases that are approximately mutually unbiased.} Interestingly, in the case of (incomplete subset of) design-structured measurements \cite{ivonovic1981,wootters1989,klappenecker2004,pittenger2004,renes2004sicpovm,gour2014,kalev2014mum,ketterer2020design,rastegin2020designpovm,yoshida2022sicpovm,rastegin2023frame}, e.g., measurements in MUBs \cite{ivonovic1981,wootters1989,klappenecker2004,pittenger2004} and mutually unbiased measurements (MUMs) \cite{kalev2014mum}, strong entropic lower bounds  \cite{ivonovic1981,wootters1989,larsen1990,sanchez1995,sanchez-ruiz1998,klappenecker2004,pittenger2004,renes2004sicpovm,wu2009,rastegin2013,gour2014,kalev2014mum,chen2015,rastegin2015,ketterer2020design,rastegin2020designpovm,huang2021,yoshida2022sicpovm,rastegin2023frame} for multiple measurements can be derived directly from upper bounds on the respective total indexes of coincidence (IC) of outcome probability distributions. Now, a meaningful question arises as to whether improved EURs valid beyond the special cases of design-structured measurements can be formulated in a similar way.

 We emphasize that, as is pointed out in \cite{huang2022,rotundo2023}, EURs like \myref{qmu} are thus far formulated in a restricted form considering that they  describe lower bounds on simply entropy sums, whereas the most general entropic way of expressing the uncertainty principle should be
 \begin{align}
 	\sum_\theta w_\theta H(\mc{M}_\theta)\geq q, \label{weur}
 \end{align}
 with $\{w_\theta\}$ being arbitrary positive weights. Conceptually speaking, requiring the weights $\{w_\theta\}$ to be equal is unnecessary since the l.h.s. of \myref{weur} well captures the presence of uncertainty regardless of the weights, i.e., it is positive for observables with no common eigenstate. Realizing this,  weighted EURs (WEURs) for multiple measurements in terms of the collision entropy as well as for two projective measurements in terms of the Shannon entropy have been established, respectively, in \mycites{huang2022} and \cite{rotundo2023}.

 In this paper, inspired by a recent work \cite{huang2022} on complementarity relations, we obtain upper bounds \eqref{icbound} on the weighted sum of IC over multiple outcome probability distributions induced by general measurements. We establish lower bounds on the weighted sum of R\'{e}nyi entropies \cite{renyi1961} for multiple generalized measurements, i.e., positive-operator-valued measures (POVMs). Compared with previous EURs, our WEURs are generally stronger and apply to versatile measurement scenarios.

This paper is structured as follows. In \mysec{pre}, we introduce general upper bounds on the IC of probability distributions induced by performing general sets of measurements on quantum systems.  In  \mysec{eur}, we propose WEURs for multiple measurements assigned with positive weights. In \mysec{application}, we take the steering test as an example to show numerically that our WEURs are advantageous in practical quantum tasks. Finally, we draw a brief conclusion in \mysec{conclusion}.

	\section{Preliminary}\label{pre}

 Each quantum measurement is described by a set of positive semi-definite operators (POVM effects) $\mc{M}=\{M_i\}$ that satisfy  $\sum_iM_i=\mathbbm{1}$, with $\mathbbm{1}$ being the identity operator.  For example, the POVM description of measuring a nondegenerate observable consists of rank-1 projectors onto its eigenbasis, called rank-1 projective measurements. Throughout the rest of this article, we frequently consider the measurement scenario where an observer chooses, according to the value of a classical random variable $\theta$ sampled from some probability distribution $\{w_\theta\}$ ($\sum_\theta w_\theta=1$, $w_\theta>0$), to perform one of a set of measurements $\{\mc{M}_\theta\}$ on individual copies of a quantum system. 	We {\tred denote by $p_{i|\theta}$  the probability of obtaining the $i$th outcome} when performing the $\theta$th measurement $\mc{M}_\theta=\{M_{i|\theta}\}_i$ on the state $\rho$, which is $p_{i|\theta}=\text{Tr}(M_{i|\theta}\rho)$ according to Born's rule.

In \mycite{huang2022}, the authors proposed an upper bound on the average information gain on quantum systems in individual trials of measurements,
\begin{align}
	\sum_{i,\theta}w_\theta\big[p_{i|\theta}-\text{Tr}(M_{i|\theta})/d\big]^2\leq \lVert\hat{g}\rVert\cdot I_{\rm com}(\rho). 
	\label{gainsum}
\end{align}
Here, $I_{\rm com}(\rho)=\text{Tr}(\rho^2)-1/d$ is the operationally invariant measure of complete information content contained in $d$-dimensional quantum states \cite{brukner1999}. $\lVert\hat{g}\rVert$ denotes the largest eigenvalue of the weighted average of view operators $\hat{g}=\sum_\theta w_\theta\hat{G}(\mc{M}_\theta)$  \cite{huang2022} (see also Appendix. \ref{appie}), which {\tred is state-independent and} depends only on the measurement scenario. In fact, the weighted average information gain given as the l.h.s. of \myref{gainsum} quantifies how much the state $\rho$ can be discriminated from the completely mixed state through the respective outcome statistics $\{p_{i|\theta}\}$ and $\{\text{Tr}(M_{i|\theta})/d\}$. The r.h.s. of \myref{gainsum}, on the other hand, limits one's ability to gain information about quantum systems in different measurement scenarios. Interestingly, \myref{gainsum} naturally explains the origin of wave-particle duality in two-way interferometers as exclusion relations between information gains in complementary measurements \cite{huang2022}.

Next, we will restrict our focus to measurements that are described by POVM effects with equal trace (ETE-POVMs). For clarity, we say a set of $l$-outcome  POVMs $\{\mc{M}_\theta\}$ are ETE-POVMs iff $\text{Tr}(M_{0|\theta})=\cdots=\text{Tr}(M_{l-1|\theta})$ for all $\theta$. Typical examples of ETE-POVMs include rank-1 projective measurements and design-structured POVMs. Notably, when considering ETE-POVMs, the l.h.s. of \myref{gainsum} essentially measures the average distance between the uniform distribution $[\text{Tr}(M_{0|\theta})/d,\cdots,\text{Tr}(M_{l-1|\theta})/d]$ which represents the maximum uncertainty (unpredictability) and the probability distributions $(p_{0|\theta},\cdots, p_{l-1|\theta})$ of measurement outcomes. In other words, \myref{gainsum} characterizes an observer's certainty about (ability to predict) outcomes of ETE-POVMs.

To establish WEURs for $l$-outcome ETE-POVMs from the certainty relation \eqref{gainsum}, let us cast \myref{gainsum} into an equivalent inequality for convenience,
\begin{align}
	\sum_{i,\theta}w_\theta p_{i|\theta}^2
	\leq 1/l+\lVert\hat{g}\rVert\cdot I_{\rm com}(\rho)=:\bar{c}.
	\label{icbound}
\end{align}
It is worth mentioning that \myref{icbound} becomes a tight equality for an arbitrary complete set of design-structured measurements with equal weights. Moreover, $\frac{1}{\Theta}\leq\lVert\hat{g}\rVert\leq1$ holds for a number $\Theta$ of rank-1 projective measurements, regardless of the weights $\{w_\theta\}$ \cite{huang2022}. In particular,  $\lVert\hat{g}\rVert=\frac{1}{\Theta}$ is saturated by random measurements in one of $\Theta$ MUBs, and for arbitrary nondegenerate observables with one or more common eigenstates there is $\lVert\hat{g}\rVert=1$. Following \mycite{huang2022}, we will call 
\begin{align}
\mc{X}_{\rm tot}=\Theta-\Big\lVert\sum_\theta\hat{G}({\mc{M}_\theta})\Big\rVert=\Theta-\lVert\hat{G}_{\rm tot}\rVert
\label{exclusivity}
\end{align}
the total measurement exclusivity, which takes value in the range $\mc{X}_{\rm tot}\in [0,\Theta-1]$ for rank-1 projective measurements. The exclusivity may be interpreted as a complement of overlap between multiple measurements, which tends to be larger for measurement bases that are less overlapped (closer to being mutually unbiased). 
 
\section{Weighted entropic uncertainty relations} \label{eur}

The R\'{e}nyi entropies are generalizations of the Shannon entropy defined as below \cite{renyi1961}
\begin{align}
	H_\alpha(\vec{p})=\frac{1}{1-\alpha}\log \sum_ip_i^\alpha, \nonumber
\end{align}
where $\vec{p}=(p_0,\ p_1,\ \cdots)$ can be any probability distribution and the parameter $\alpha>0$ and $\alpha\neq1$. The Shannon entropy $H(\vec{p})=-\sum_ip_i\log p_i=\lim_{\alpha\to1}H_\alpha(\vec{p})$ is thus recovered {\tred in the limit} $\alpha\to1$. {\tred  R\'{e}nyi entropies has many essential significance in cryptography and information theory. As a noteworthy example,  the minimum entropy $H_\infty(\vec{p})=-\log p_{\rm max}$ $(p_{\rm max}=\max_i\{p_i\})$ characterizes the number of random bits that can be extracted from a random variable obeying the distribution $\vec{p}$. For more discussions on the basic properties and applications of R\'{e}nyi entropies, we recommend the review \cite{coles2017}. }

\begin{figure}[b]
	\centering
	\includegraphics[width=8.6cm]{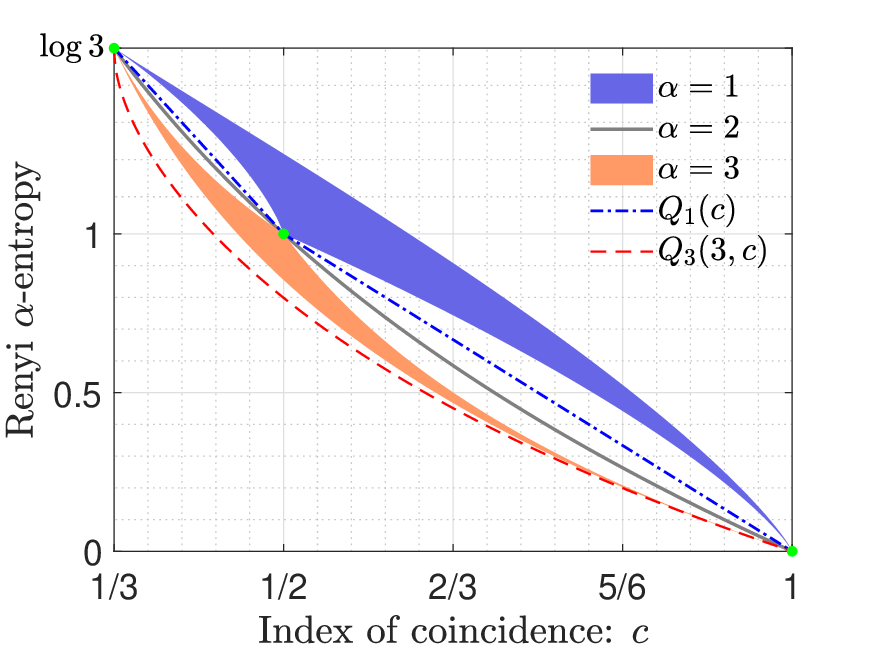}
   \caption{The IC-entropy diagrams and convex estimations   $Q_1$ (\myref{Q1};  dot-dashed blue line) and $Q_3$ (\myref{Q2}; dashed red line) of their lower boundaries, respectively for the Shannon entropy (blue region) and the R\'{e}nyi 3-entropy (orange region).}
	\label{entropydiagram}
\end{figure}

Our discussions on WEURs revolve around the relationship between the IC of probability vectors and the corresponding R\'{e}nyi entropies. The IC of a probability vector $\vec{p}$ refers to the probability that two independent random variables drawn from $\vec{p}$ take the same value, that is,  $c(\vec{p})=\sum_{i}p_i^2$.  Obviously, $c(\vec{p})\in[1/l,1]$ if the length of $\vec{p}$ is $l$.  Observe that $H_\alpha(\vec{p})=0$ iff $c(\vec{p})=1$ (e.g., $p_0=1, p_1=\cdots=p_{l-1}=0$), and $H_\alpha(\vec{p})=\log l$ iff $c(\vec{p})=1/l$ (i.e., $p_0=\cdots=p_{l-1}=1/l$).  When  $l=3$, the IC-entropy diagrams---the ranges of the map $\vec{p}\rightarrow\big(c(\vec{p}), H_\alpha(\vec{p})\big)$---are plotted in \myfig{entropydiagram} for  $\alpha=1,2,3$ respectively. We can see they intersects at the green points $\{(1/3,\log 3), (1/2,1), (1,0)\}$ on the curve $(c,-\log c)$ of the R\'{e}nyi 2-entropy. Similar results hold also for general $l\geq2$. This is because $H_\alpha(\vec{p})$ is monotonic decreasing of $\alpha$ except when the nonzero probabilities of $\vec{p}$ is uniform, say $p_0=\cdots=p_{n-1}=1/n$ for some integer $n\in[1,l]$,  then  $c(\vec{p})=1/n$ and $H_\alpha(\vec{p})=\log n$ is independent of $\alpha$.

Let us first consider the case $\alpha\geq2$. To estimate the lower boundary of the IC-entropy diagram we adopt the function  \cite{huang2021} 
\begin{align}
Q_\alpha(l,c)=\frac{\alpha\log p_a}{1-\alpha}+
\frac{\log l \times\log\Big[1+(l-1)^{\frac{2}{\alpha}}\frac{p_b^2}{p_a^2}\Big]}{(1-\alpha)\log[1+(l-1)^{\frac{2}{\alpha}}]},
\label{Q2}
\end{align}
where $p_a=\frac{1+\sqrt{(lc-1)(l-1)}}{l}$ and $p_b=\frac{1-\sqrt{(lc-1)/(l-1) }}{l}$. This estimation function (see the dashed red line in \myfig{entropydiagram} for the case $\alpha=3$) is optimal when $\alpha=2$ or $  +\infty$, as well as when  $l=2$, and it remains a good estimation in other situations, especially when $l$ is large. Additionally, \myref{Q2} is convex with respect to $c$ (see a detailed proof in Appendix. \ref{app_Q}), combined with the IC bound \eqref{icbound}  we immediately have the theorem below.

 \emph{Theorem 1.}  Suppose $\{\mc{M}_\theta\}$ are $l$-outcome ET-POVMs to be performed on the state $\rho$ with selection probabilities $\{w_\theta\}$, and $\hat{g}$ is the average view operator. When $\alpha\geq2$ the average R\'{e}nyi $\alpha$-entropy satisfies
\begin{align}
\sum_\theta w_\theta H_\alpha(\mc{M}_\theta)_\rho \geq Q_\alpha\left[l,1/l+\lVert\hat{g}\rVert\cdot I_{\rm com}(\rho)\right]=: q_\alpha.     \label{alphaentropy}
\end{align}
 Interestingly, the uncertainty lower bound given as the r.h.s. of  \myref{alphaentropy} decreases monotonically with the quantity $\lVert\hat{g}\rVert\cdot I_{\rm com}(\rho)$. {\tred Meanwhile, for fixed measurements and selection probabilities, $q_\alpha$ reaches its minimum at all pure states, thus becoming state-independent in the sense that it is an uncertainty lower bound valid for all quantum states in the Hilbert space considered.}

 \emph{Corollary 1.} For rank-1 projective measurements on {\tred individual qubits in the state $\rho$}
 \begin{align}
     q_2= -\log\left[\frac{1}{2}+\lVert\hat{g}\rVert\cdot I_{\rm com}(\rho)\right]\geq\log\left[\frac{2}{1+\lVert\hat{g}\rVert}\right].
     \label{2entropy}
 \end{align}

\begin{figure}[t]
	\centering
	\includegraphics[width=8.6cm]{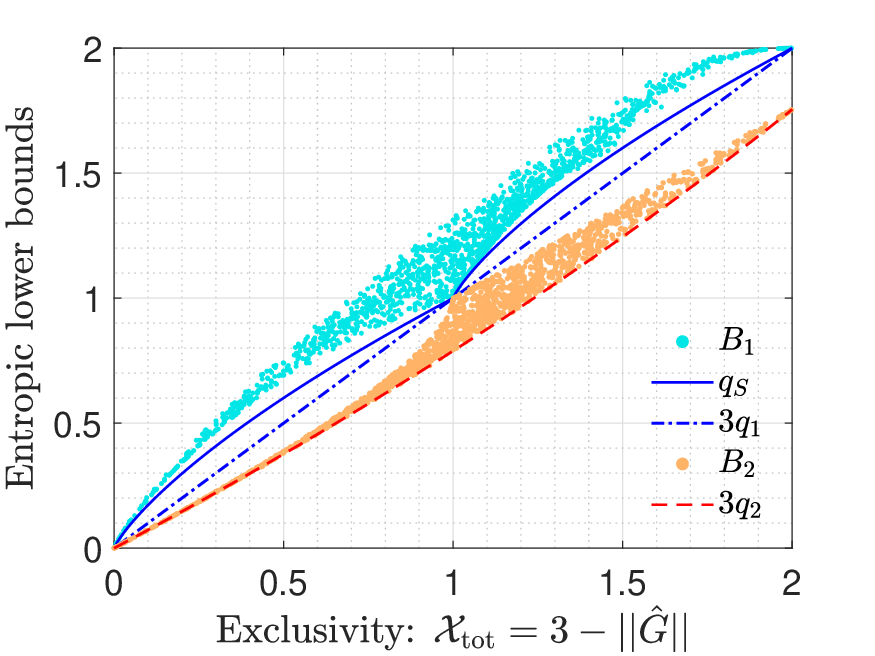}
	\caption{State-independent lower bounds on the sum of Shannon entropies {\tred ($q_S$  \eqref{qubitentropy}, real blue line; $q_1$ \eqref{qubitentropy1}, dot-dashed blue line) and R\'{e}nyi 2-entropies ($q_2$ \eqref{2entropy}, dashed red line) } for 1600 randomly generated sets of three single-qubit observables with equal weights: a comparison with the respective numerical optimal bounds $B_1$ (blue dots) and $B_2$ (orange dots).}
	\label{entropy}
\end{figure}

Equation \eqref{2entropy} is compared in \myfig{entropy} with the corresponding numerical optimal entropic lower bound for three nondegenerate observables with equal weights. As shown, $q_2$ is very strong, especially when the exclusivity is around $\mc{X}_{\rm tot}=2$  (MUBs) or $\mc{X}_{\rm tot}=0$ (compatible bases).

As for the Shannon entropy, the corresponding lower boundary of the IC-entropy diagram consists of concave curves each joining a pair of neighbor points with coordinates on the set $\{({\tred1/n},\log n)|n=1,\cdots,l\}$  \cite{harremoes2001,huang2021}  (see the green points in \myfig{entropydiagram}). Substituting these curves for line sections we arrive at the estimation function  (see the dot-dashed blue line in \myfig{entropydiagram})
\begin{align}
    Q_1(c)=\log n-(n+1)(nc-1)\log (1+1/n).
    \label{Q1}
\end{align}
Here $n=\lfloor1/c\rfloor$ denotes the round-down of $1/c$ to the nearest integer. Similar to the estimation function \eqref{Q2} for $\alpha\geq2$, equation \eqref{Q1} is convex with respect to $c$. Combining  \myref{Q1} with the IC bound \eqref{icbound}, the following theorem is then obvious.

 \emph{Theorem 2.}  Suppose $\{\mc{M}_\theta\}$ are $l$-outcome ETE-POVMs to be performed on the state $\rho$ with selection probabilities $\{w_\theta\}$, and $\hat{g}$ is the associated average view operator. The average Shannon entropy satisfies
	\begin{align}
		\sum_{\theta}w_\theta H(\mc{M}_\theta)_\rho\geq Q_1\left[1/l+\lVert\hat{g}\rVert\cdot I_{\rm com}(\rho)\right]=:q_1.  \label{sentropyline}
	\end{align}

\emph{Corollary 2.} For rank-1 projective measurements on {\tred individual qubits in the state $\rho$}
\begin{align}
		q_1= 1 -2\lVert\hat{g}\rVert\cdot I_{\rm com}(\rho)\geq1-\lVert\hat{g}\rVert.
\label{qubitentropy1}
\end{align}

If the inverse of $1/l+\lVert\hat{g}\rVert\cdot I_{\rm com}(\rho)$ happens to be an integer, say $n$,  then  $q_1=\log n$ would be the best uncertainty bound that can be obtained from the IC bound  \eqref{icbound}. But this is not the case in general. Luckily, when restricted to ETE-POVMs with equal weights, we can make full use of  \myref{icbound} to derive improved EURs.

\emph{Theorem 3.} Suppose $\{\mc{M}_\theta\}_{\theta=1}^\Theta$ is a set of $l$-outcome ETE-POVMs to be performed on {\tred individual  quantum systems in} the state $\rho$. Then, the sum of Shannon entropies satisfies
	\begin{align}
		\sum_{\theta}H(\mc{M}_\theta)_\rho&\geq k\log (n-1)+(\Theta-k-1)\log n\nonumber\\
		-&(1-p)\log [(1-p)/(n-1)]-p\log p=: q_S.  \label{sentropy}
	\end{align}
{\tred Here,} with $\bar{c}= 1/l+\frac{1}{\Theta}\lVert \hat{G}_{\rm tot}\rVert \cdot I_{\rm com}(\rho)$ being the average IC, $n=\lceil1/\bar{c}\rceil$ and $k=\lfloor n(n-1)(\bar{c}-1/n)\Theta\rfloor$. {\tred In addition, } $p\in[0,\frac{1}{n}]$ is the solution to
	$(1-p)^2\frac{1}{n-1}+p^2= \Theta\bar{c}-\frac{k}{n-1}-\frac{\Theta-k-1}{n}$. We refer to Appendix. \ref{appeur} for a detailed proof of Theorem 3.

{\tred For a complete set of $d+1$ MUBs (CMUBs) in $d$-dimensional Hilbert space, note that the corresponding total view operator must satisfy $\lVert\hat{G}_{\rm tot}\rVert=1$ (see also Appendix. \ref{appie}). Substituting $I_{\rm com}(\rho)\leq 1-1/d$ into \myref{sentropy} then leads us to the strong entropic bound previously obtained in \mycites{ivanovic1992, sanchez1993, sanchez1995}.
	\begin{align*}
		q_S\geq\left\{\begin{aligned}
			&(d+1)\log \frac{d+1}{2} \hspace{2.9cm}\text{odd $d$},\\
			&\frac{d}{2}\log \frac{d}{2}+\Big(\frac{d}{2}+1\Big)\log \Big(\frac{d}{2}+1\Big)\hspace{0.74cm} \text{even $d$}.
		\end{aligned}\right.
	\end{align*}
	This immediately indicates that \myref{sentropy} would be strong for approximately CMUBs as $q_S$ is continuous with respect to measurements.
}

\emph{Corollary 3.} For {\tred $\Theta$} rank-1 projective measurements {\tred on individual qubits in the state $\rho$}
\begin{align}
		q_S=&h_{\rm bin}\Big(\frac{1}{2}+\frac{1}{2}\sqrt{2\lVert \hat{G}_{\rm tot}\rVert I_{\rm com}(\rho) -k}\Big)+\Theta-1-k\nonumber\\
	\geq& h_{\rm bin}\Big(\frac{1}{2}+\frac{1}{2}s\Big)+\Theta-1-\lfloor\lVert \hat{G}_{\rm tot}\rVert \rfloor,
 \label{qubitentropy}
	\end{align}
where $h_{\rm bin}(p)=-p\log p-(1-p)\log (1-p)$, $k=\lfloor2\lVert \hat{G}_{\rm tot}\rVert\cdot I_{\rm com}(\rho) \rfloor$, and  $s=(\lVert \hat{G}_{\rm tot}\rVert-\lfloor \lVert \hat{G}_{\rm tot}\rVert\rfloor)^{1/2}$.

{\tred Both $q_1$ \eqref{qubitentropy1} and $q_S$ \eqref{sentropy} decreases monotonically with the operationally invariant information $I_{\rm com}(\rho)$ contained in the quantum systems to be measured, and achieve their state-independent minimum at pure states. For two rank-1 projective measurements $(\Theta=2)$}, \myref{qubitentropy} reduces to $q_S\geq h_{\rm bin}(\frac{1}{2}+\frac{1}{2}\sqrt{2c_{\rm max}-1})$.  This recovers an earlier bound reported in \mycite{sanchez-ruiz1998}, which is known to be tighter than $q_{\rm MU}$ \eqref{qmu}.  For rank-1 projective measurements onto three MUBs $({\tred \Theta=3,}\ \lVert\hat{G}_{\rm tot}\rVert=1)$,  $q_S=2+S(\rho)$ is known to be tight \cite{coles2011}, where $S(\rho)=-\text{Tr}(\rho\log\rho)$ is the von Neumann entropy. We emphasize here that \myref{sentropy} is a general result valid beyond the aforementioned simple cases.   In \myfig{entropy}, the state-independent forms of $q_1$ and $q_S$ are compared with the respective numerical optimal uncertainty bounds for three random nondegenerate observables of qubits. As depicted, both of them are tight for MUBs ($\mc{X}_{\rm tot}=2$) and remain to be strong when $\mc{X}_{\rm tot}<2$.


\begin{figure*}
	\centering
	\subfigure[]{\label{fig3-a}
		\includegraphics[width=8.6cm]{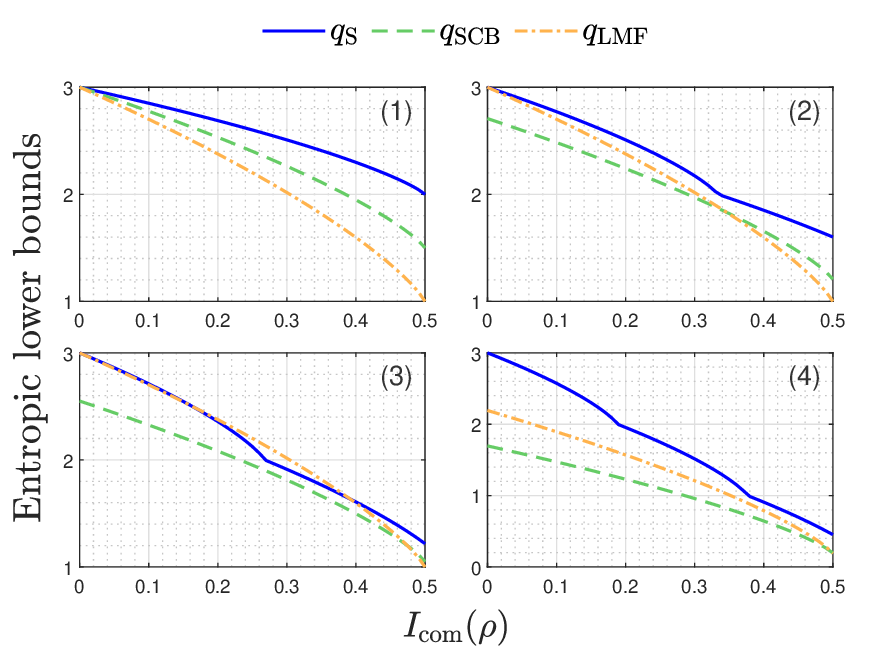}}
	\hspace{0.005\linewidth}
	\subfigure[]{\label{fig3-b}
		\includegraphics[width=8.6cm]{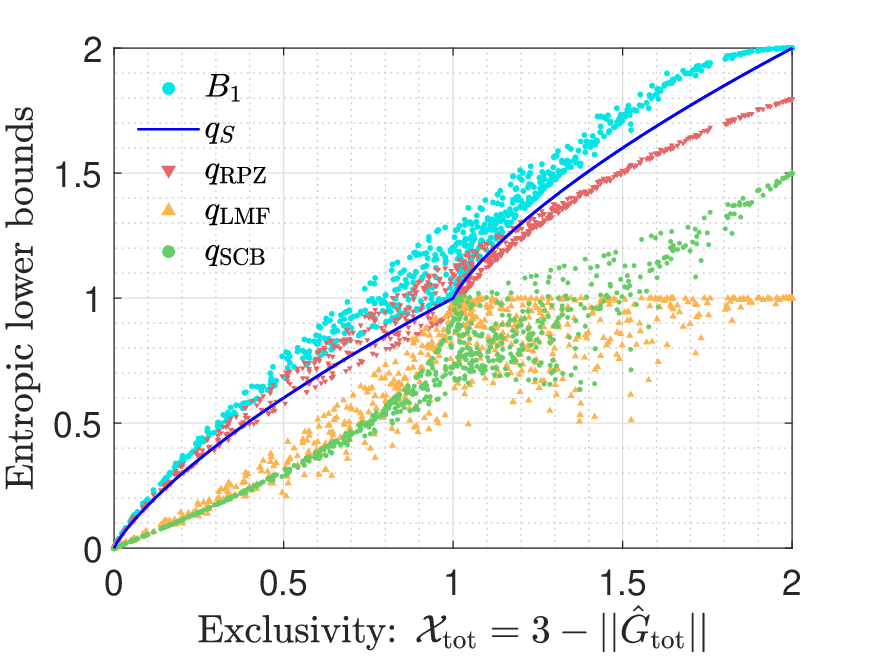}}
	\hspace{0.005\linewidth}
	\vfill
	\subfigure[]{\label{fig3-c}
		\includegraphics[width=8.6cm]{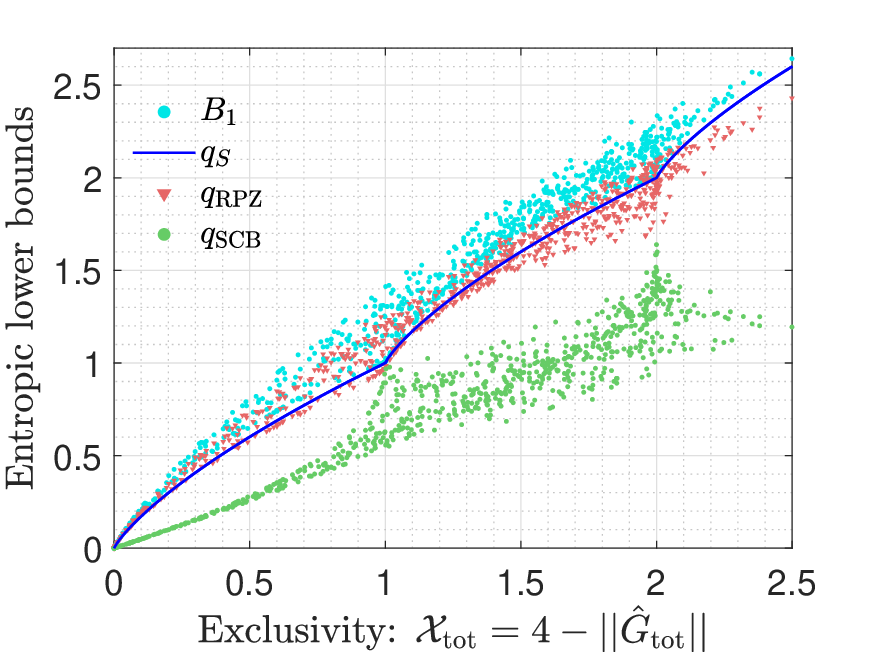}}
	\hspace{0.005\linewidth}
	\subfigure[]{\label{fig3-d}
		\includegraphics[width=8.6cm]{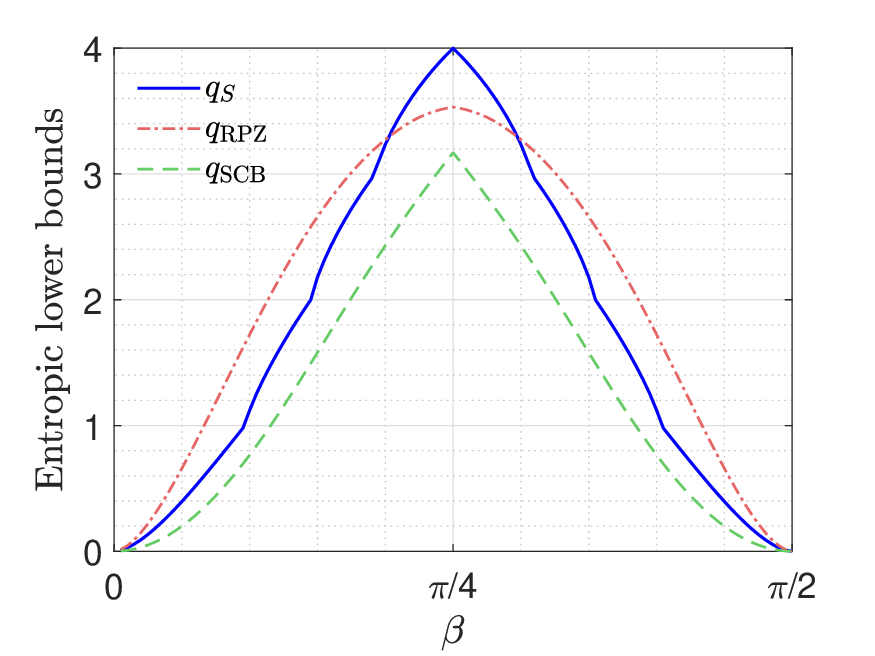}}
	\hspace{0.005\linewidth}
    	\caption{\tred{Lower bounds on the Shannon entropy sums. $(a)$ Comparisons between $q_S$ \eqref{qubitentropy}, $q_{\rm SCB}$ \eqref{qXie} and $q_{\rm LMF}$ \eqref{qLiu} for three single-qubit observables:  (1) \{$\sigma_x$, $\sigma_y$, $\sigma_z$\}. (2) \{$\sigma_x$, $\sigma_y$,  $\frac{1}{2}\sigma_x+\frac{\sqrt{3}}{2}\sigma_z$\}. (3) \{$\sigma_x$, $\sigma_y$,  $\frac{\sqrt{3}}{2}\sigma_x+\frac{1}{2}\sigma_z$\}. (4) \{$\sigma_x$, $\frac{\sqrt{3}}{2}\sigma_x+\frac{1}{2}\sigma_y$,  $\frac{\sqrt{3}}{2}\sigma_x+\frac{1}{2}\sigma_z$\}. $(b)$ Comparison between the state-independent forms of bounds $q_S$, $q_{\rm LMF}$, $q_{\rm SCB}$, $q_{\rm RPZ}$ and the numerical optimal bound $B_1$ for 750 randomly generated sets of three bases in $\mc{H}_2$. $(c)$ Comparison between state-independent entropic bounds for 750 randomly generated sets of four bases in $\mc{H}_2$. $(d)$ Entropic bounds for four bases in $\mc{H}_3$ defined by \myref{4bases3}.}
    \label{fig3}}
\end{figure*}

\section{Comparison of bounds}

To further {\tred compare our EURs with} those ones obtained in previous works, we remark that inequalities (\ref{alphaentropy}, \ref{sentropyline}, \ref{sentropy}) extend a series of  EURs \cite{ivonovic1981,wootters1989,larsen1990,sanchez1995,sanchez-ruiz1998,klappenecker2004,pittenger2004,renes2004sicpovm,wu2009,rastegin2013,gour2014,kalev2014mum,chen2015,rastegin2015,ketterer2020design,rastegin2020designpovm,huang2021,yoshida2022sicpovm,rastegin2023frame} for (incomplete) design-structured measurements to WEURs for general ETE-POVMs.  Therefore, we only need to consider those EURs that hold for measurements without special structure. {\tred Liu \emph{et al.} generalized the two-bases bound $q_{\rm MU}$ \eqref{qmu} to multiple bases as below\cite{liu2015}
\begin{align}
	q_{\rm LMF}=-\log b+(\Theta-1)S(\rho),
 \label{qLiu}
\end{align}
where $c_{\rm max}^{(\theta,\theta')}=\max_{i,j}\{c^{(\theta,\theta')}_{i,j}\}$ denotes the maximal overlap between the $\theta$th and $\theta'$th bases and
\begin{align}
	b=\max_{i_\Theta}\Big\{\sum_{i_2,\cdots,i_{\Theta-1}}\max_{i_1}\{c^{(1,2)}_{i_1,i_2}\}\prod_{\theta=2}^{\Theta-1}c^{(\theta,\theta+1)}_{i_\theta,i_{\theta+1}}\Big\}.\nonumber
\end{align}
Notably, $q_{\rm LMF}$ \eqref{qLiu} is never weaker than $q_{\rm MU}$ for any pair of bases, whereas it is not tight for multiple MUBs.  As is pointed out by Xie \emph{et al.} \cite{xie2021},  the entropic lower bound 
\begin{align}
	q_{\rm SCB}=-\frac{1}{\Theta-1}\sum_{\theta>\theta'}\log c_{\rm max}^{(\theta,\theta')}+\frac{\Theta}{2}S(\rho)
	\label{qXie}
\end{align}
simply constructed from $q_{\rm MU}$ by considering the bases pairwisly can be stronger than $q_{\rm LMF}$ when the bases are approximately mutually unbiased.

We present in \myfig{fig3-a} numerical comparisons between $q_S$ \eqref{qubitentropy}, $q_{\rm LMF}$ \eqref{qLiu} and the simply constructed bound $q_{\rm SCB}$ \eqref{qXie} on the sum of Shannon entropies over three single-qubit observables. As shown, $q_S\geq q_{\rm SCB}$ holds for all the four sets of observables considered. Additionally, although $q_{\rm LMF}$ is relatively weaker} for pure states [$S(\rho)=0; I_{\rm com}(\rho)=0.5$], it can be stronger than $q_{\rm SCB}$ and $q_S$ for mixed states. It is worth noting that $q_{\rm LMF}>q_S$ may hold for mixed states in the special case when two of the three observables are approximately compatible while the third one is complementary to them [see \myfig{fig3-d}-3].  But more generally, $q_{\rm LMF}$ tends to be weaker than $q_S$, especially for observables that are close to being mutually complementary or commutable [see \myfig{fig3-d}-(1, 2, 4)].

{\tred In fact, for $\Theta$  MUBs in $\mc{H}_d$, the state-independent forms of Eqs.  (\ref{qLiu}, \ref{qXie}) are $q_{\rm LMF}=\log d$ and $q_{\rm SCB}=\frac{\Theta}{2}\log d$. When $\Theta\geq\sqrt{d}+1$, $q_S\geq \Theta q_1\geq \Theta q_2=\Theta\log\frac{d\Theta}{d+\Theta-1}\geq \Theta\log\sqrt{d}=q_{\rm 	SCB}\geq q_{\rm LMF}$.  This already demonstrates that Eqs. (\ref{sentropyline}, \ref{sentropy}) would be stronger than $q_{\rm LMF}$ and $q_{\rm SCB}$, at least for $\Theta\geq\sqrt{d}$ +1 bases in $\mc{H}_d$ that are sufficiently close to being mutually unbiased. In \myfig{fig3-b}, we move on to take into consideration the famous state-independent bound  $q_{\rm RPZ}$ obtained from the direct sum majorization relations by Rudnicki \emph{et al.} \cite{rudnicki2014}. As depicted, both $q_S$ and  $q_{\rm RPZ}$  are never weaker than the state-independent forms of $q_{\rm LMF}$ and $q_{\rm SCB}$ for arbitrary three bases in $\mc{H}_2$. Moreover, $q_S$ attains the optimal bound $B_1$ for three MUBs in $\mc{H}_2$ ($\mc{X}_{\rm tot}=2$), $q_{\rm LMF}=1<q_{\rm SCB}=1.5<q_{\rm RPZ}\approx1.8<q_S=2$ , and remains to be stronger than $q_{\rm RPZ}$ for approximately MUBs. On the other hand, for bases that are close enough to be compatible ($\mc{X}_{\rm tot}\lesssim 1$), $q_{\rm RPZ}$ is stronger in general.

When considering four bases in $\mc{H}_2$, we can see from \myfig{fig3-c} that $q_S>q_{\rm RPZ}$ tends to hold for measurement bases with large exclusivity and, roughly speaking, $q_{\rm RPZ}>q_S$ holds for bases that are close enough to be compatible. In \myfig{fig3-d}, we make a further comparison between $q_S$ and the bounds $q_{\rm RPZ}$ and $q_{\rm SCB}$ for $\Theta=4$ bases in $\mc{H}_3$, including the computational basis and three unitary transformations of it below,
\begin{align}
	U_1=F^{4\beta/\pi},\ U_2=E(F)^{4\beta/\pi},\ U_3=E^2(F)^{4\beta/\pi}. 
	\label{4bases3}
\end{align}
Here $E$ is diagonal with eigenvalues $\{1, e^{i2\pi  /3}, e^{i2\pi  /3}\}$, and $F$ denotes the discrete Fourier transform, i.e, $F_{jk}=\frac{1}{\sqrt{3}}e^{i2\pi jk/3}$. Again, $q_S=4$ is tight \cite{riccardi2017} for MUBs $(\beta=\pi/4)$, in which case it is stronger than $q_{\rm RPZ}$. 
}

 \section{Entropic steering criterion}\label{application}

{\tred In this section we discuss the applications of our EURs in steering detection. The concept of steering \cite{uola2020, guhne2023} dates back to Einstein, Podolskey, and Rosen's \cite{EPR1935} prominent observation that quantum correlation allows one to predict the outcome of measuring one particle based on the measurement performed on a distant particle. One possible explanation for this kind of correlation between distant measurement outcomes is the local hidden state (LHS) model. To illustrate, consider a bipartite state shared between Alice and Bob, then Bob's local state after Alice's measurement on subsystem A is said to admit a LHS model if it can be decomposed as
 \begin{align}
 	\sigma_{a|\theta}=\sum_\lambda\pi(\lambda)p_A(a|\theta,\lambda)\rho_\lambda, 
 	\label{lhs}
 \end{align}
where $\lambda$ denotes the value of an assumed hidden variable subject to some probability distribution $\pi(\lambda)$, $p_A(a|\theta,\lambda)$ denotes Alice's probability to obtain the $a$th outcome when she chooses to perform the measurement labeled by $\theta$, and $\{\rho_\lambda\}$ are local hidden states on Bob's side independent of Alice's measurements. Thus, \myref{lhs} essentially describes a particular form of correlation between Bob's local states $\{\sigma_{a|\theta}\}$ and Alice's measurement outcomes $\{(a,\ \theta)\}$. What's interesting is quantum states can be steerable and violate steering inequalities \cite{cavalcanti2009} such that they do not admit a LHS model. In Schro\"{o}dinger's words, ``the steering forces Bob to believe that Alice can influence his particle from a distance'' \cite{schrodinger1935}.  From a modern point of view, steering signifies the presence of entanglement, but not necessarily Bell nonlocality \cite{uola2020}.

Steering has many applications in quantum information processing (see the reviews \cite{uola2020, guhne2023} and references therein for details), wherein a key point is to detect when and how much a certain steering inequality can be violated, or Bob's ability to predict his local measurement outcomes conditioned on Alice's measurement results. To this end, we utilize the conditional R\'{e}nyi entropy \cite{iwamoto2014} to quantify Bob's uncertainty about (inability to predict) his local measurement outcomes
\begin{align*}
	H_\alpha=\frac{\alpha}{1-\alpha}\log\Big[\sum_{j}p_A(j)\lVert p_{B|A}(i|j)\rVert_\alpha\Big],
\end{align*}
where $p_{B|A}(i|j)$ is the conditional probability and $\lVert\cdot\rVert$ denotes the $\alpha$-norm. As a straightforward generalization of the entropic steering criterion \cite{krivachy2018,ketterer2020design}, if Alice is unable to convince Bob that the state $\rho_{AB}$ is entangled by performing the local measurements $\{\mc{M}^A_\theta\}$ on subsystem A we have
\begin{align}
	\sum_\theta w_\theta H_\alpha(\mc{M}^B_\theta|\mc{M}^A_\theta)_{\rho_{AB}}\geq q_\alpha (\{\mc{M}^B_\theta, w_\theta\}).
	\label{steerineq}
\end{align}
Here $q_\alpha (\{\mc{M}^B_\theta, w_\theta\})$ denotes the state-independent entropic bounds given as the r.h.s. of Eqs. (\ref{alphaentropy}, \ref{sentropyline}) when Bob chooses to perform the measurement $\mc{M}^B_\theta$ with probability $w_\theta$ on subsystem B.  Violation of \myref{steerineq} necessarily implies Bob's local uncertainty can be reduced given Alice's measurement results, so that the state $\rho_{AB}$ is steerable from Alice to Bob.

It is worth noting that to accomplish a steering task Alice must choose proper measurements, that is, nonjointly measurable (incompatible)  measurements \cite{busch1996,heinosaari2016}.}  Joint measurability is an operationally motivated extension of the commutativity of observables to generalized measurements.  White noise robustness is a commonly used measure of incompatibility. It refers to the critical value $\eta^*$ of $\eta$ $(0\leq\eta\leq1)$ below which a set of noisy measurements $\{\eta E_{i|\theta}+(1-\eta)\frac{1}{d}\mathbbm{1}_d\}$ become jointly measurable ($\eta=0$ corresponds to trivial measurements, which are always compatible).  Next we study how much white noise could corrupt the incompatibility of Alice's measurements to keep inequality \eqref{steerineq} saturated. Following \mycites{ketterer2020design,krivachy2018}, we exploit the equivalence \cite{uola2015one,uola2014joint} between incompatibility and steering of a maximally entangled state, {\tred that is, whenever Alice chooses incompatible measurements she will succeed in convincing Bob that the state shared between them is entangled. }


 \begin{figure}[t]
	\includegraphics[width=8.6cm]{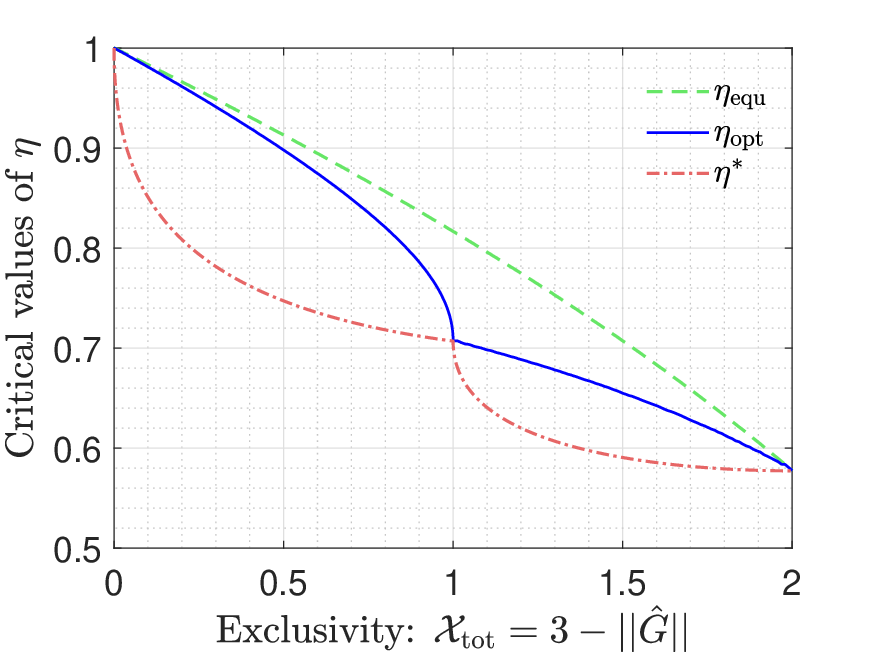}
	\caption{Estimations of  the incompatibility of three qubit-observables based on \myref{steerineq}  when $\alpha=+\infty$ (minimum entropy). For three complementary observables $(\mc{X}_{\rm tot}=2)$,  \myref{steerineq} yields $\eta_{\rm equ}\approx 0.577$  when the weights $w_1=w_2=w_3$ are equal, which well estimates the corresponding incompatibility $\eta^*=\frac{1}{\sqrt{3}}$. More generally, optimizing the weights leads to a better estimation $\eta_{\rm opt}$ and  $\eta^*\leq\eta_{\rm opt}<\eta_{\rm equ}$. }
	\label{incomp}
\end{figure}

As a simle example, we then utilize \myref{steerineq} to estimate the incompatibility of single-qubit observables.
Suppose now the maximally entangled state $\rho_{ AB}=\frac{1}{\sqrt{2}}(\ket{00}-\ket{11})$ is shared between Alice and Bob, and in each round of the test, if Alice chooses to measure the observable $\sigma$, Bob measures its noisy counterpart $\eta\sigma+(1-\eta)\frac{1}{2}\mathbbm{1}_2$.  For simplicity, we consider three observables in the form  $\{\cos\beta_1\sigma_x+\sin\beta_1\sigma_z,\cos\beta_2\sigma_y+\sin\beta_2\sigma_z,\sigma_z\}$, with $\beta_1=0$ and $0\leq\beta_2<\frac{\pi}{2}$ or  $\beta_2=\frac{\pi}{2}$ and $0<\beta_1\leq\frac{\pi}{2}$. We present  in \myfig{incomp} some  numerical results about the threshold value $\eta_{\rm equ}$ ($\eta_{\rm opt}$) of $\eta$  for Bob's measurements with equal (optimal) weights to violate \myref{steerineq}.  As shown, \myref{steerineq} yields $\eta_{\rm equ}\approx0.577$ for the case of three complementary observables $(\beta_1=\beta_2=0$, $\mc{X}_{\rm tot}$=2). This value coincides with the noise robustness $\eta^*=\frac{1}{\sqrt{3}}$. {\tred More generally we have $\eta_{\rm equ}>\eta_{\rm opt}$, therefore optimizing the weights in \myref{steerineq}} leads to superior performance. {\tred We would like to remark that the performance of \myref{steerineq} has been underestimated in the preceding analysis, as we restricted Alice's measurement choice to be the noiseless counterpart of Bob's measurement for simplicity.}

\section{Conclusion}\label{conclusion}

In this work, we have derived R\'{e}nyi EURs for multiple measurements assigned with positive weights from upper bound \eqref{icbound} on the respective IC of outcome probability distributions. On one hand, our results extend a series of independent investigations \cite{ivonovic1981,wootters1989,larsen1990,sanchez1995,sanchez-ruiz1998,klappenecker2004,pittenger2004,renes2004sicpovm,wu2009,rastegin2013,gour2014,kalev2014mum,chen2015,rastegin2015,ketterer2020design,rastegin2020designpovm,huang2021,yoshida2022sicpovm,rastegin2023frame} on EURs for design-structured measurements to WEURs for much more general measurements. On the other hand, we verified both analytically and numerically that our Shannon EURs are generally stronger than the generalizations{\tred, $q_{\rm LMF}$ \eqref{qLiu} and $q_{\rm SCB}$ \eqref{qXie}, }of Maassen and Uffink's famous EUR \eqref{qmu} to multiple bases. {\tred Our bound $q_S$ \eqref{sentropy} can also outperform the strong direct-sum majorization bound $q_{\rm RPZ}$ \cite{rudnicki2014} for certain measurements, especially for approximately mutually unbiased bases (MUBs). }
Taking the steering test as an example, we also demonstrated numerically that WEURs could achieve better performance in practical applications simply by optimizing the weights assigned to different measurements. Crucially, the optimization process can be readily accomplished on a classical computer without incurring any additional quantum costs.

Our investigation provides new insights into the interpretation of quantum uncertainty from an entropic perspective and we expect it to inspire future in-depth researches on the applications of WEURs in quantum information theory.
To improve our results, future works will take into consideration other eigenvalues of the average view operator to tighten the IC bound \eqref{icbound}. Exploring new applications of WEURs also presents an intriguing avenue for future investigation.

\acknowledgments

This work is supported by the National Natural Science Foundation of China (No. 12175104 and No. 12274223),  the National Key Research and development Program of China (No. 2023YFC2205802), the Innovation Program for Quantum Science and Technology (No. 2021ZD0301701), the Natural Science Foundation of Jiangsu Province (No. BK20211145), the Fundamental Research Funds for the Central Universities (No. 020414380182), the Key Research and Development Program of Nanjing Jiangbei New Area (No. ZDYD20210101),  the Program for Innovative Talents and Entrepreneurs in Jiangsu (No. JSSCRC2021484), and the Program of Song Shan Laboratory (Included in the management of Major Science and Technology Program of Henan Province) (No. 221100210800-02).

\appendix

\section{View operator}\label{appie}
The view operator of a measurement $\mc{M}=\{M_i\}$ on $d$-dimensional systems is defined as below \cite{huang2022} 
\begin{align}
	\hat{G}(\mc{M})=\sum_i\hat{G}(M_i)=\sum_i|\tilde{M}_i\rangle\langle \tilde{M}_i|,
	\label{viewoperator}
\end{align}
where {\tred $\tilde{M}_i= M_i-\frac{1}{d}{\rm Tr}(M_i) \mathbbm{1}_d$ is traceless and $\ket{\tilde{M}_{i}}=\sqrt{d}(\tilde{M}_{i}\otimes\mathbbm{1}_d)\ket{\psi_d}$} is a vector on the product space $\mc{H}_d\otimes\mc{H}_d$ which is orthogonal to the maximally entangled isotropic state $\ket{\psi_d}=\frac{1}{\sqrt{d}}\sum_{i=0}^{d-1}\ket{i}\otimes\ket{i}^*$. 

We list here some basic properties of view operators (see detailed proofs in \mycite{huang2022}).\\
 (1) View operators are positive semi-definite,  $\hat{G}\geq0$ on the $(d^2-1)$-dimensional subspace $\mc{H}_{\perp\psi_d}$ of $\mc{H}_d\otimes\mc{H}_d$ that consists of vectors orthogonal to $\ket{\psi_d}$.\\
   (2) View operators vanish for POVMs whose effects are all proportional to the identity operator $\mathbbm{1}_d$ on $\mc{H}_d$, i.e., POVMs satisfying  $M_i=\frac{1}{d}\text{Tr}(M_i)\mathbbm{1}_d$ for each $i$.\\
     (3) For any rank-1 projective measurement, the associated view operator is an identity operator on a $(d-1)$-dimensional subspace of $\mc{H}_{\perp\psi_d}$.\\
      (4) The view operators of measurements in MUBs are mutually orthogonal and the combined view operator $\hat{G}_{\rm tot}=\sum_{\theta=1}^{d+1}\hat{G}(\mc{M}_\theta)$ associated with $d+1$ MUBs constitute the identity operator on $\mc{H}_{\perp\psi_d}$. \\
        (5) The largest eigenvalue of the view operator associated with an ETE-POVM $\{M_i\}$ equals to the second largest eigenvalue of the overlap matrix $W_{i,j}=\text{Tr}(M_iM_j)$. \\
          (6) Let $\hat{G}_{\rm tot}$ denote the combined view operator associated with a set of $\Theta$ orthonormal bases of $\mc{H}_d$, then $1\leq\lVert\hat{G}_{\rm tot}\rVert\leq \Theta$. Moreover,  $\lVert\hat{G}_{\rm tot}\rVert=\Theta$ is achieved by measurements with which the corresponding overlap matrix, defined as $W_{i|\theta,j|{\theta'}}=\text{Tr}(M_{i|\theta}M_{j|\theta'})$, is reducible, e.g., observables with a common eigenstate. \\
              (7) Let $\hat{g}=\sum_\theta w_\theta\hat{G}(\mc{M}_\theta)$ denote the average view operator associated with a set of $\Theta$ orthonormal bases in $\mc{H}_d$, then $\lVert\hat{g}\rVert\leq \sum_\theta w_\theta\lVert\hat{G}(\mc{M}_\theta)\rVert=\sum_\theta w_\theta=1$ and $\lVert\hat{g}\rVert\geq \max_\theta \{w_\theta\lVert\hat{G}(\mc{M}_\theta)\rVert\}\geq\max_\theta\{w_\theta\}\geq\frac{1}{\Theta}$. 

\section{Lower boundary of IC-entropy diagram}\label{app_Q}
Here we show that $Q_\alpha(l,c)$ \eqref{Q2} are convex estimations of lower boundaries of the IC-R\'{e}nyi entropy diagrams. In \mycites{harremoes2001, huang2021}, it has been shown that when $\alpha\geq2$ the lower boundary of each IC-R\'{e}nyi entropy diagram is saturated by probability distributions in the form 
\begin{align}
	\vec{p}=(p_a,\ p_b, \cdots, p_b),
\end{align}
where $\{p_a=\frac{1+\sqrt{(lc-1)(l-1)}}{l},\ p_b=\frac{1-\sqrt{(lc-1)/(l-1) }}{l}\}$  is the solution to the equations $p_a+(l-1)p_b=1$ and $p_a^2+(l-1)p_b^2=c$. 

First, we need to show $H_\alpha(\vec{p})=\frac{1}{1-\alpha}\log[p_a^\alpha+(l-1)p_b^\alpha]\geq Q_\alpha(l,c)$. Observe that
	\begin{align}
		&(1-\alpha)[H_\alpha(\vec{p})-Q_\alpha(l,c)]\nonumber\\
		=&\log\left[1+(l-1)\left(\frac{p_b}{p_a}\right)^\alpha\right]\nonumber\\
		&\hspace{3em}-\frac{\log l \times\log\left[1+(l-1)^{2/\alpha}\left(\frac{p_b}{p_a}\right)^2\right]}{\log\left[1+(l-1)^{2/\alpha}\right]}.
		\label{ent-esti}
	\end{align}
	Let $z=p_b/p_a\in[0,1]$, then the r.h.s. of \myref{ent-esti} shares the same sign as the following term
	\begin{align}
		Z=\frac{\log\left[1+z^\alpha(l-1)\right]}{\log\left[1+z^2(l-1)^{2/\alpha}\right]}-\frac{\log l}{\log\left[1+(l-1)^{2/\alpha}\right]}.
	\end{align}
	Let $0\leq\zeta(z)=z^2(l-1)^{\frac{2}{\alpha}}\leq(l-1)^{2/\alpha}$. $Z$ is non-increasing with respect to $\zeta(z)$ when $\alpha\geq2$, which immediately leads to
	\begin{align}
		Z=\frac{\log\left[1+\zeta^\frac{\alpha}{2}(z)\right]}{\log\left[1+\zeta(z)\right]}-\frac{\log\left[1+\zeta^\frac{\alpha}{2}(1)\right]}{\log\left[1+\zeta(1)\right]}\leq0.
	\end{align}
	Therefore $H_\alpha(\vec{p})\geq Q_\alpha(l,c)$.

	We proceed to show $Q_\alpha(l,c)$ is convex with respect to $c$ when $\alpha\geq2$. Let us begin with the function below 
	\begin{align}
		(\alpha-1)Q_\alpha=&\left[\alpha-\frac{2\log l}{\log[1+(l-1)^{2/\alpha}]}\right]A\nonumber\\
		&\hspace{1.5cm}+\frac{\log l}{\log[1+(l-1)^{2/\alpha}]} B,
	\end{align}	
	where $A=-\log p_a$, $B=-\log\left[p_a^2+(l-1)^{\frac{2}{\alpha}}p_b^2\right]$. The coefficient of $B$ is obviously positive. As for $A$, 
	\begin{align}
		\alpha-\frac{2\log l}{\log[1+(l-1)^{\frac{2}{\alpha}}]}=\alpha-\frac{\alpha\log l}{\log[1+(l-1)^{\frac{2}{\alpha}}]^{\frac{\alpha}{2}}}\geq0
	\end{align}
	is non-negative too. Next we show $A$ and $B$ are convex with respect to $c$, respectively. Differentiate the equations $p_a+(l-1)p_b=1$ and $p_a^2+(l-1)p_b^2=c$ leads us to
	\begin{align}
		\frac{dp_a}{dc}=\frac{1}{2(p_a-p_b)},\ \frac{dp_b}{dc}=-\frac{1}{2(l-1)(p_a-p_b)}. 
		\label{tool}
	\end{align}
    Based on \myref{tool} we have
	\begin{equation}\label{der_1}
		\frac{dA}{dc}=-\frac{1}{p_a}\frac{dp_a}{dc}=\frac{-1}{2p_a(p_a-p_b)}<0.
	\end{equation}
	Considering that $p_a+(l-1)p_b=1$, now we know that $p_a$ increases with $c$ and $p_b$ decreases with $c$. Consequently, $p_a-p_b$ increases with $c$. According to \myref{tool} we then have $d^2p_a/dc^2<0$ and 
	\begin{equation}
		\frac{d^2A}{dc^2}=\frac{1}{p_a^2}\left(\frac{dp_a}{dc}\right)^2-\frac{1}{p_a}\frac{d^2p_a}{dc^2}>0.
		\label{d2A}
	\end{equation}
	The above proves $A$ is convex with respect to $c$. When $\alpha\geq 2$, according to \myref{tool} we have 
	\begin{align}
		&\frac{dB}{dc}=\frac{-2}{2^{-B}\ln 2}\left[p_a\frac{dp_a}{dc}+(l-1)^{\frac{2}{\alpha}}p_b\frac{dp_b}{dc}\right]  \nonumber\\
		=&\frac{-1}{2^{-B}\ln2}\frac{p_a-(l-1)^{(2/\alpha-1)}p_b}{p_a-p_b}\nonumber\\
		=&\frac{-1}{2^{-B}\ln2}\left[1+\frac{1-(l-1)^{(2/\alpha-1)}}{p_a-p_b}p_b\right]\leq0.
	\end{align}
	$B$ is non-increasing means $\frac{-1}{2^{-B}}$ is non-decreasing, while the term in the square braket is obviously non-increasing due to the monotonicity of $p_b$ and $p_a-p_b$. Thus $d^2B/dc^2>0$, which, combined with \eqref{d2A}, completes the proof that $Q_\alpha$ is convex with respect to $c$.

\section{Shannon EURs}\label{appeur}

To prove Theorem 3, here we mainly utilize the method for constructing Shannon EURs from IC introduced in \mycite{huang2021}. Some intermediate  conclusions in the first subsection have been obtained also in \mycites{sanchez1993,harremoes2001}.

\subsection{Shannon lower bound for a single distribution}\label{sd}
Consider all possible probability distributions with index of coincidence (IC) being $c$, namely, $\sum_ip_i^2=c$, it was shown in \mycites{sanchez1993,harremoes2001,chen2015,huang2021} that the probability vector $\vec{p}$ that minimizes the Shannon entropy takes the form
\begin{align}
&\vec{p}=(\overbrace{p_a,\cdots,p_a, p_b}^{n\ \text{in total}}), \ p_a\geq p_b=1-(n-1)p_a\geq0;  \label{cond1}\\
&c(\vec{p})=(n-1)p_a^2+p_b^2=c. \label{cond2}
\end{align}
$\vec{p}$ is a linear combination of two uniform distributions of length $n$ and $n-1$ respectively, and is fully determined by the above equations:   \myref{cond1} requires $\frac{1}{n}\leq c\leq \frac{1}{n-1}$,  namely, $n=\lceil1/c\rceil$. Immediately $p_a=\frac{1}{n}+\frac{1}{n}\sqrt{(cn-1)/(n-1)}$ and $p_b=\frac{1}{n}-\frac{1}{n}\sqrt{(cn-1)(n-1)}$ is determined by \myref{cond2}.

\begin{figure}[b]
	\includegraphics[width=8.6cm]{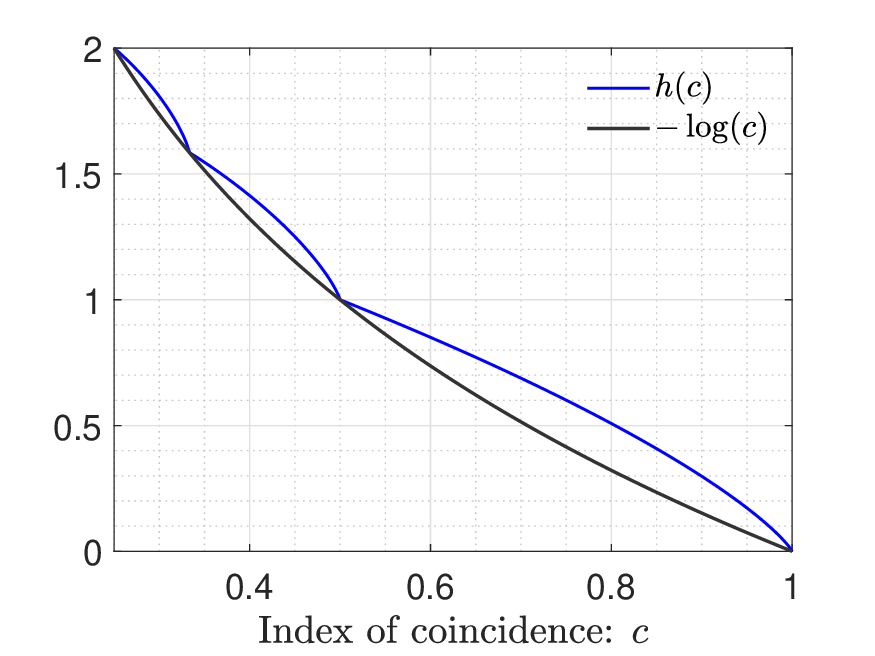}
	\caption{The diagram of $h(c)$ for $c\in[1/4,1]$.}    \label{ic-entropy}
\end{figure}

So the Shannon entropy of any probability distribution with IC being $c$ is lower bounded by
\begin{align}
h(c)=-(n-1)p_a\log p_a-p_b\log p_b,
\end{align}
with
\begin{align}
n&=\lceil1/c\rceil,\ p_a=\frac{1}{n}+\frac{1}{n}\sqrt{(cn-1)/(n-1)},\\
p_b&=\frac{1}{n}-\frac{1}{n}\sqrt{(cn-1)(n-1)}.
\end{align}
The diagram of $h(c)$ is plotted in \myfig{ic-entropy} for $c\in[1/4,1]$.

There are some useful \textbf{properties} of $h(c)$. (i) $h(c)$ is decreases monotonically with $c$ \cite{sanchez1993,harremoes2001,chen2015,huang2021}. (ii) For any positive integer $k$, $h(c)$ is concave with respect to $c\in [1/(k+1),1/k]$ \cite{harremoes2001,huang2021}.  (iii) If $1/c=k$ is an integer, then $n=k$, $p_a=p_b=1/k$, and $h(c)=-\log k$ \cite{harremoes2001,huang2021}. (4) For arbitrary two positive integers $k_1>k_2\geq1$ there is \cite{huang2021}
\begin{align}
&h(1/k_1)+h(1/k_2+s)\geq h(1/k_1+s)+h(1/k_2)\label{prop1}\\
&\text{for}\ 0\leq s\leq \frac{1}{k_1-1}-\frac{1}{k_1},\nonumber\\
&h(1/k_1-s)+h(1/k_2)\geq h(1/k_1)+h(1/k_2-s)\label{prop2}\\
&\text{for}\ 0\leq s\leq \frac{1}{k_1}-\frac{1}{k_1+1}.\nonumber
\end{align}
The properties above is enough to derive the best lower bound on the Shannon entropy for multiple probability distributions based only on an upper bound on the average IC (see the next subsection). We present below the detailed proof of property (4).

The proof is straight forward, for $c\in [\frac{1}{n}, \frac{1}{n-1})$, let $s=c-\frac{1}{n}$ we have
\begin{widetext}

\begin{align}
&h(c)=A(s,n)\nonumber\\
=&\log n-\frac{n-1}{n}\left(1+\sqrt{\frac{ns}{n-1}}\right)\log \left(1+\sqrt{\frac{ns}{n-1}}\right)
-\frac{1}{n}\left(1-\sqrt{sn(n-1)}\right)\log\left(1-\sqrt{sn(n-1)}\right) .
\end{align}
To show \myref{prop1} we only need to show $B(s,n)=\log n-A(s,n)$ is an increasing function of $n$ for $s\in [0,\frac{1}{n(n-1)}]$,

\begin{align}
\frac{\partial B(s,n)}{\partial n}=\frac{\frac{sn}{2}+\sqrt{sn(n-1)}}{n^2\sqrt{sn(n-1)}}\left[\log \frac{1+\sqrt{sn/(n-1)}}{1-\sqrt{sn(n-1)}} -\frac{sn^2}{\frac{sn}{2}+\sqrt{sn(n-1)}}\right].
\end{align}

Let $0\leq u=\sqrt{sn(n-1)}<1$, the term in square brackets is
\begin{align}
&\log \frac{n-1+u}{(n-1)(1-u)}-\frac{u^2n}{\frac{u^2}{2}+u(n-1)}
\geq\log \frac{n-1+u}{(n-1)(1-u)}-\frac{un}{(n-1)}, \label{test1}
\end{align}
The r.h.s. of \myref{test1} increases monotonically with $u$, obviously it's nonnegative. To show \myref{prop2} we only need to show $C(s,n)=A\left(\frac{1}{n(n-1)}-s,n\right)-\log (n-1)$ is an increasing function of $n$ for $s\in [0,\frac{1}{n(n-1)}]$.

\begin{align}
C(s,n)=&\log \frac{n}{n-1}-\frac{1}{n}\left(n-1+\sqrt{1-sn(n-1)}\right)\log \left(1+\frac{\sqrt{1-sn(n-1)}}{n-1}\right)\nonumber\\
-&\frac{1}{n}\left(1-\sqrt{1-sn(n-1)}\right)\log\left(1-\sqrt{1-sn(n-1)}\right) .
\end{align}
\begin{align}
\frac{\partial C(s,n)}{\partial n}=\frac{1+\frac{sn}{2}-\sqrt{1-sn(n-1)}}{n^2\sqrt{1-sn(n-1)}}\log \frac{1+\frac{1}{n-1}\sqrt{1-sn(n-1)}}{1-\sqrt{1-sn(n-1)}}-\frac{1-\sqrt{1-sn(n-1)}}{n(n-1)}.
\end{align}
Let $0\leq u=\sqrt{1-sn(n-1)}<1$ we have
\begin{align}
\frac{\partial C(s,n)}{\partial n}=&\frac{2n(1-u)-(1-u)^2}{2n^2(n-1)u}\log \frac{n-1+u}{(n-1)(1-u)}-\frac{1-u}{n(n-1)} \nonumber\\
=&\frac{2n(1-u)-(1-u)^2}{2n^2(n-1)u}\left[\log \frac{n-1+u}{(n-1)(1-u)}-\frac{2nu}{2n-(1-u)}\right] \label{test2}.
\end{align}

Similar to \myref{test1}, the term in square brackets of \myref{test2} is positive
\begin{align}
\log \frac{n-1+u}{(n-1)(1-u)}-\frac{2nu}{2n-(1-u)}\geq\log \frac{n-1+u}{(n-1)(1-u)}-\frac{2nu}{2n-1}\geq 0.
\end{align}

\end{widetext}

\subsection{The sum of Shannon entropies over multiple probability distributions}
Suppose  $c_{\rm tot}\geq\sum_{\theta=1}^\Theta c(\vec{p}_\theta)$ is an upper bound on the total IC of $\Theta$ $(\Theta\geq2)$ length-$l$ probability vectors $\{\vec{p}_\theta\}$. Then, the best lower bound on the sum of Shannon entropies that can be obtained from $c_{\rm tot}$ is
\begin{align}
h_\Theta(l,c_{\rm tot})=\min\limits_{\sum_{\theta=1}^\Theta c(\vec{p}_\theta)=c_{\rm tot}}\ \sum_\theta h[c(\vec{p}_\theta)]. \label{seur}
\end{align}
To evaluate $h_\Theta(l,c_{\rm tot})$, we first examine what the steepest-descent method leads us to, and then prove it is indeed optimal instead of a local minimum.

Let $i\times U_j$ denote $i$ uniform distributions of length $j$ and $I_\Theta(l,k)=\frac{\Theta-k}{l}+\frac{k}{l-1}$ denote the total IC of the following $\Theta$ probability distributions: $k\times U_{l-1}$ $(k=0,1,\cdots,\Theta-1)$ and $(\Theta-k)\times U_l$. When $c_{\rm tot}=\Theta/l=I_M(l,0)$, obviously $h_\Theta(l,c_{\rm tot})=\Theta\log l$. As $c_{\rm tot}$ increases, the steepest descent of entropy (SDE) is $D_\Theta(l,c_{\rm tot})=(\Theta-1)\log l+h(c_{\rm tot}-\frac{\Theta-1}{l})$ when $c_{\rm tot}<I_\Theta(l,1)=(\Theta-1)/l+1/(l-1)$. This is because $h(c)$ is concave with respect to $c$ for $c\in [1/l,1/(l-1)]$. Similarly, for $I_\Theta(l,1)\leq c_{\rm tot}\leq I_\Theta(l,2)=(\Theta-2)/l+2/(l-1)$,  according to \myref{prop1} the SDE leads us to $h_\Theta(l,c_{\rm tot})=(\Theta-2)\log l+\log (l-1)+h(c_{\rm tot}-\frac{\Theta-2}{l}-\frac{1}{l-1})$. Furthermore, for general $c_{\rm tot}$:
\begin{widetext}
\begin{align}
D_\Theta(l,c_{\rm tot})=\left\{
\begin{aligned}
	&(\Theta-1)\log l+h[c_{\rm tot}-(\Theta-1)/l],\ I_\Theta(l,0)\leq c_{\rm tot}<I_\Theta(l,1)\\
	&(\Theta-2)\log l+\log (l-1)+h[c_{\rm tot}-I_\Theta(l,1)+1/l],\  I_\Theta(l,1)\leq c_{\rm tot}<I_\Theta(l,2)\\
	&\hspace{5cm}\vdots\\
	&(\Theta-k-1)\log l+k\log (l-1)+h(c_{\rm tot}-I_\Theta(l,k)+1/l),\  I_\Theta(l,k)\leq c_{\rm tot}< I_\Theta(l,k+1)\\
	&\hspace{5cm}\vdots\\
	&(\Theta-k-1)\log n+k\log (n-1)+h(c_{\rm tot}-I_\Theta(n,k)+1/n),\  I_\Theta(n,k)\leq c_{\rm tot}< I_\Theta(n,k+1)
\end{aligned}
\right.
\end{align}
\end{widetext}
Observe in the last equality above $\Theta/n\leq I_\Theta(n,k)\leq c_{\rm tot}< I_\Theta(n,k+1)\leq \Theta/(r-1)$, which implies $n=\lceil \Theta/c_{\rm tot}\rceil$. Further, since
$$\frac{1}{n}\leq c_{\rm tot}-I_\Theta(n,k)+\frac{1}{n}=c_{\rm tot}-\frac{\Theta-1}{n}-\frac{k}{n(n-1)}< \frac{1}{n-1},$$
or equivalently
$$n(n-1)(c_{\rm tot}-\frac{\Theta}{n})-1< k\leq n(n-1)(c_{\rm tot}-\frac{\Theta}{n}),$$
we have $k=\lfloor n(n-1)(c_{\rm tot}-\frac{\Theta}{n})\rfloor$. Therefore the steepest-descent method leads us to
\begin{align}
D_\Theta(l,c_{\rm tot})=&(\Theta-k-1)\log n+k\log (n-1) \nonumber\\
+&h(c_{\rm tot}-\frac{\Theta-k-1}{n}-\frac{k}{n-1}),  \label{optbound}
\end{align}
with $n=\lceil \Theta/c_{\rm tot}\rceil$ and $k=\lfloor n(n-1)(c_{\rm tot}-\frac{\Theta}{n})\rfloor$. Ignoring zero-valued probabilities, \myref{optbound} is saturated by the following $\Theta$ probability distributions: $(\Theta-k-1)\times U_n$, $k\times U_{n-1}$ and one distribution that is a linear combination of $U_n$ and $U_{n-1}$.

To see  why \myref{optbound} is indeed the optimal lower bound, i.e., $D_\Theta(l,c_{\rm tot})=h_\Theta(l,c_{\rm tot})$, observe first the property (ii) of $h(c)$ ensures that the set of nonnegative numbers $\{c_1,\cdots,c_\Theta\}$ satisfying $\sum_\theta c_\theta=c_{\rm tot}$ and $h_\Theta(l,c_{\rm tot})=\sum_\theta h(c_\theta)$ contains at most one element that is not an inverse of an integer. Without losing generality, suppose $c_1=\frac{1}{k_1},\cdots, c_{\Theta-1}=\frac{1}{k_{\Theta-1}}$ are inverse of integers satisfying $k_1\geq k_2\cdots \geq k_{\Theta-1}\geq1$ and $k_1>1$. Note that \myref{prop2} requires $k_1-k_{\Theta-1}\leq1$, otherwise $\log k_1+\log k_{\Theta-1}\geq \log(k_1-1)+h(\frac{1}{k_{\Theta-1}}-\frac{1}{k_1-1}+\frac{1}{k_1})$ contradicts the assumption that $\sum_\theta h(c_\theta)=h_\Theta(l,c_{\rm tot})$; \myref{prop2} also requires $c_\Theta\geq\frac{1}{k_{\Theta-1}+1}$, or $h(c_\Theta)+\log k_{\Theta-1}\geq -\log \lfloor 1/c_\Theta\rfloor+h\left(1/k_{\Theta-1}-1/\lfloor 1/c_\Theta\rfloor+c_\Theta\right)$ contradicts $\sum_\theta h(c_\theta)=h_\Theta(l,c_{\rm tot})$. Moreover, \myref{prop1} requires $c_\Theta\leq \frac{1}{k_{\Theta-1}}$ if $k_1-k_{\Theta-1}=1$, otherwise $\log k_1+h(c_\Theta)\geq h \left(1/k_1-1/\lceil1/c_\Theta\rceil+c_\Theta\right)-\log(\lceil1/c_\Theta\rceil)$  contradicts $\sum_\theta h(c_\theta)=h_\Theta(l,c_{\rm tot})$. The above restrictions on $\{c_1,\cdots,c_\Theta\}$ promises that \myref{optbound} is the best entropic lower bound that can be obtained based only on an upper bound on IC.

\bibliographystyle{apsrev4-2-title}	
\bibliography{WEUR240114}

\end{document}